\documentstyle[epsf,twoside]{ptptex}
\notypesetlogo  
\title{The $D_s(2317)$ and $D_s(2463)$ Mesons as Scalar and Axial-Vector Chiralons in the\\
       Covariant Level-Classification Scheme
}
\author{%
Shin {\sc Ishida}$^1$\footnote{Associate member.}, Muneyuki Ishida$^2$, Toshihiko Komada$^3$,
Tomohito Maeda$^1$,\\
Masuho Oda$^4$, Kenji Yamada$^3$ and Ichiro Yamauchi$^5$
}
\inst{%
$^{1}$ Research Institute of Quantum Science, 
College of Science and Technology\\
Nihon University, Tokyo 101-0062, Japan\\
$^{2}$ Department of Physics, Meisei University, Tokyo 191-8506. Japan\\
$^{3}$ Department of Engineering Science, Junior College Funabashi Campus,\\
 Nihon University, Funabashi 274-8501, Japan\\
$^{4}$ Faculty of Engineering, Kokushikan University, Tokyo 154-8515, Japan\\
$^{5}$ Department of Mechanical Engineering, Tokyo Metropolitan College of Technology,
 Tokyo 140-0011, Japan
}
\recdate{%
\today
}
\abst{%
The new narrow mesons observed recently in the final states $D_s^+ \pi^0$ and $D_s^{*+} \pi^0$
are pointed out to be naturally assigned as the ground-state scalar and axial-vector chiralons
in the $(c\bar s)$ system, which would newly appear in the covariant hadron-classification scheme
proposed a few years ago.
}

\begin{document}

\maketitle
\setcounter{tocdepth}{4}

\section{Introduction}

{\it ($Covariant$ $Classification$ $Scheme$ $and$ $Chiral$ $states/Chiralons$)}\ \ \ \ 
A few years ago we have proposed a covariant level-classification scheme\cite{rf2}
for hadrons, unifying the seemingly contradictory two, non-relativistic and extremely relativistic, 
viewpoints. (Its essential points are reviewed in our review articles\cite{rf1}.)
Here the framework is manifestly Lorentz-covariant and the space for the static symmetry is 
extended from that of non-relativistic (NR) scheme to
\begin{eqnarray}
SU(6)_{SF} & \bigotimes & \begin{array}{|c|} \hline SU(2)_\rho \\   \hline \end{array}\  
{\scriptstyle \bigotimes} \ O(3)_L , 
\label{eq2}
\end{eqnarray}
where a new additional $SU(2)$-space for the $\rho$-spin ($\rho$- and $\sigma$- spin being 
Pauli-matrices in the decomposition of Dirac $\gamma$ matrices:
$\gamma \equiv \sigma {\scriptstyle \bigotimes} \rho$) is introduced for covariant 
description\footnote{
It is to be noted that in our scheme the \underline{squared-mass} spectra are globally
$\tilde U(12)$-symmetric\cite{rf2} and the mass spectra themselves are able to be reconciled with the broken
chiral symmetry.
}
of hadron spin-wave function (WF).

The spin WF for the quark-antiquark meson systems are generally given by the 
Bargmann-Wigner (BW) spinors, and are represented as the bi-Dirac spinors
$W_\alpha^\beta  =  u_\alpha \bar v^\beta$ \ $:$\ \  $\alpha =(\rho_3,\sigma_3)$,\  
$\beta =(\bar\rho_3,\bar\sigma_3)$,
where $\alpha$($\beta$) denotes the suffices of Dirac spinors of quarks(anti-quarks)  
represented by the eigenvalues of $\rho$-spin and $\sigma$-spin,
and $\bar\rho_3\equiv -\rho_3^T$.
In the light-quark (LL) meson system the states
 with $(\rho_3,\bar\rho_3)=(+,-),(-,+),(-,-)$
(to be called ``Chiralons"), in addition to the states with 
with $(\rho_3,\bar\rho_3)=(+,+)$
(to be called ``Paulons"),
 are expected to be realized in nature.\footnote{
The name of Paulon$/$Chiralon is reflecting that the covariant spin
WF of the state with $(\rho_3,\bar\rho_3)=(+,+)$ is equivalent to that of 
the non-relativistic Pauli-WF and the states appear also in the 
non-relativistic scheme, while the states with the other values of 
$(\rho_3,\bar\rho_3)$ will be expected to appear newly in the covariant scheme
due to the chiral symmetry. 
}
The $\sigma$ meson is naturally assigned as a ground state chiralon in the LL meson system.
In the heavy-light (HL) meson system the states with
$(\rho_3,\bar\rho_3) =  (+,+)$ and 
$(+,-)$ are expected to be 
realized, reflecting the physical situation that the HL meson system has 
the non-relativistic
$SU(6)_s$ spin symmetry (the relativistic, chiral symmetry) concerning the constituent 
Heavy quarks (Light quarks), and is predicted to exist the scalar and axial-vector chiralons with
$(\rho_3,\bar\rho_3)=(+,-)$, as well as the pseudo-scalar and vector Paulons with $(\rho_3,\bar\rho_3)=(+,+)$.
Actually we have pointed out some experimental indications for the existence\cite{Ito} of the axial-vector chiralon
in the $(c\bar n)$ system and the scalar chiralon in the $(b\bar n)$ system.
Our relevant $D_s(2317)/D_s(2463)$ mesons are naturally assigned as 
the scalar$/$axial-vector chiralons in the $(c\bar s)$
ground states, and play the role of 
chiral partners of the already established Paulons $D_s/D_s^*$. 

{\it ($Description$ $of$ $HL$ $Mesons$)}\ \ \ \ 
The WF of HL mesons are described as
\begin{eqnarray}
\Phi_A{}^B (x,y) & \sim & \psi_{Q,A}(x) \bar\psi^{q, B} (y) \label{eq6}\\
A=(\alpha ,a),\  B=(\beta ,b);
 & & \  \alpha ,\beta =(1\sim 4);\ a=(c\ {\rm or}\ b),\ b=(u,d,s), \nonumber 
\end{eqnarray}
and are assumed to satisfy the master Klein-Gordon equation of Yukawa-type\cite{rfY}.
The squared-mass operator is assumed to contain no light-quark Dirac matrices $\gamma^{(q)}$
in the ideal limit, leading to the chiral symmetric global structure of \underline{squared-mass} spectra.
The WF is separated into the two parts, the one of plane-wave center of mass motion and 
the other of internal WF: 
The internal WF with definite total spin $J$ is expanded in terms of respective eigen functions 
$W(P)$ on spinor-space and $O(P_N,r)$ on internal space-time, 
where $W_\alpha^\beta (P_N)$ and $O(P_N,r)$ are covariant tensors respectively, in the 
$\tilde U(4)_{D.S.}$ (pseudo-unitary Dirac spinor) space and the $O(3,1)_L$ (Lorentz-space).

{\it $Spin$ $WF/BW$-$spinors$}\ \ \ \ 
As the complete set of spinor-space eigen-functions we choose the BW spinors, which are defined as solutions of the 
(local) Klein-Gordon equations.
For the HL-mesons we have the two physical solutions:
\begin{eqnarray}
U_\alpha{}^\beta (P) &\equiv& u_\alpha^{(Q)} (P) \bar v_{(\bar q)}^\beta (P);\ 
C_\alpha{}^\beta (P) \equiv u_\alpha^{(Q)} (P) \bar v_{(\bar q)}^\beta (-P),\    \label{eq11}
\end{eqnarray}
As is evidently seen from Eq.~(\ref{eq11}), 
through the chiral transformation on light anti-quarks
$\bar v(P)\gamma_5=\bar v (-P)$, the former is changed into the latter as $U(P)\gamma_5=C(P)$.
They are decomposed into the pseudo-scalars$/$vectors, and scalars$/$axial-vectors, respectively, as
\begin{eqnarray}
U_\alpha{}^\beta (v) &=& 1/2\sqrt 2\ (1-iv\cdot\gamma )\left[ i\gamma_5 P_s(P) + i\tilde\gamma_\mu V_\mu (P) \right],
\nonumber \\ 
C_\alpha{}^\beta (v) &=& 1/2\sqrt 2\ (1-iv\cdot\gamma )\left[ S(P) + i\gamma_5\tilde\gamma_\mu A_\mu (P) \right],\ \ 
(P_\mu \tilde\gamma_\mu = 0,\  v_\mu \equiv P_\mu /M).\ \ \ \ \ \   
\label{eq12}
\end{eqnarray}

{\it $Internal$ $space$-$time$ $WF/Yukawa$ $oscillators$}\ \ \ \ 
As the complete set of space-time eigen-functions we choose the covariant, 4-dimensional Yukawa
oscillator functions. By imposing the freezing relative-time condition they become effectively the conventional,
3-dimensional oscillators:
\begin{eqnarray}
\langle P_\mu r_\mu \rangle =  \langle P_\mu  p_\mu \rangle = 0 & \Rightarrow & O(3,1)_L \approx O(3)_L\ .\ \ \   
\label{eq13}
\end{eqnarray}

\section{Mass spectra for low-lying $D$ and $D_s$-mesons}

\begin{figure}[t]
\epsfysize=10cm
   \centerline{\epsffile{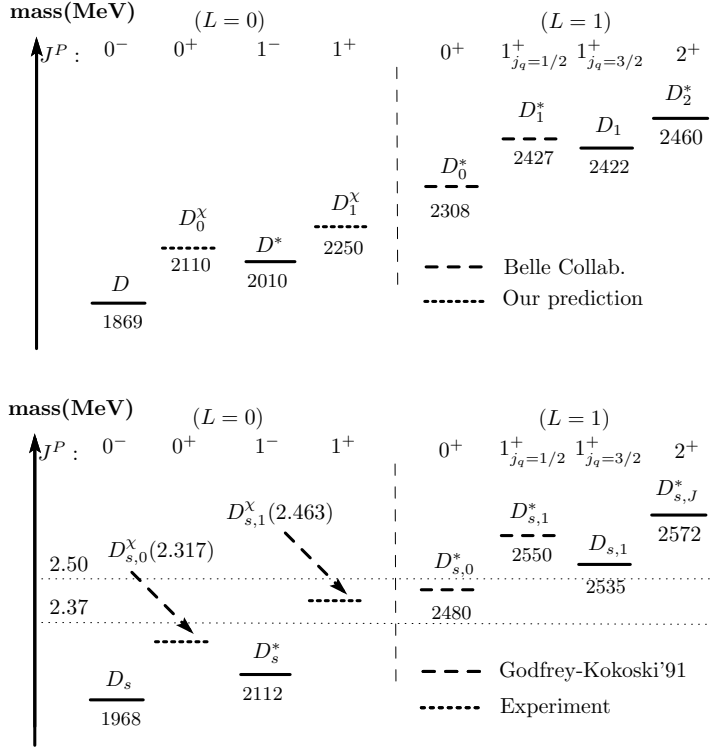}}
\caption{$D$-meson spectra(upper figure); dotted(dashed) line denotes prediction
of our theory (experimental value reported by Belle collab.\cite{Belle}). 
$D_s$ meson spectra(lower figure); dotted(dashed) line denotes experimental value
(prediction by Godfrey-Kokoski\cite{Godfrey}).
The solid line in both figures denotes the value given by PDG.\cite{rf7}  }
 \label{fig1}
\end{figure}

Since of the static symmetry (\ref{eq2}) the global mass spectra are given by
\begin{eqnarray}
M_N^2 &=& M_0^2 + N \Omega ,\  N\equiv 2n+L  ,   
\label{eq14}
\end{eqnarray}
leading to phenomenologically well-known Regge trajectories.
The masses of ground state mesons,
$P_s$, $V_\mu$, $S$ and $A_\mu$ are degenerate in the ideal limit, and 
they are expected to split with each others
between chiral partners (spin partners) 
by the spontaneous breaking of the chiral symmetry
(the perturbative QCD spin-spin interaction) as
\begin{eqnarray}
M_0 (0^- / 1^-) & \stackrel{<}{\scriptscriptstyle \sim} & M_0  (0^+ / 1^+) < M_1(L=1)\ .    
\label{eq15}
\end{eqnarray}
Actually, in \S 3, we shall apply the chiral symmetric Yukawa type interaction in non-derivative form, 
${\cal L}_{ND}$ Eq.~(\ref{eq25}), to treat the interaction of the light quarks inside of HL mesons
with the $\sigma$-meson nonet $s=s^a\lambda^a/\sqrt 2$. 
In spontaneous chiral symmetry breaking, $s$ takes the vacuum expectation value 
$\langle s \rangle \equiv diag \{a,a,b\}$, which induce
the splittings between chiral partners.
Because of the HQS,
the universal relation, 
\begin{eqnarray}
\Delta M^\chi = M_0 (0^+) - M_0  (0^-) = M_0 (1^+) - M_0  (1^-),    
\label{eq16}
\end{eqnarray}
is expected to be valid within the same light-quark-flavor mesons; and the relation of those 
between the different light-quark configurations as
$ \Delta M^\chi (c\bar n) / \Delta M^\chi (c\bar s) =a/b$.
In SU(3) linear $\sigma$ model\cite{CH,Mune}, $a$ and $b$ are related with pion and kaon decay constants as
$a\equiv f_\pi / \sqrt 2$, $(a+b)/2\equiv f_K / \sqrt 2$.
Through the experimental values of 
$\Delta M^\chi (c\bar s)=350$MeV$/c^2$ and $f_\pi /f_K$(which gives $a/b = 1/1.44$), 
we predict $\Delta M^\chi (c\bar n)=240$MeV$/c^2$.
In Fig.~\ref{fig1} (a) and (b) we show, respectively, 
the low-lying $D$ meson\footnote{
As shown in Fig. 1(a), we predict the existence of $D_0^\chi (2110)$.
New experimental data of $D\pi$ mass spectra by Belle\cite{Belle} show a peak structure with very wide width, 
which is identified as a single resonance $D_0^*(2308)$. 
We consider the possibility\cite{Yama1} that this peak structure is explained 
by the interference of two resonant states, $D_0^\chi (2110)$ 
and the $P$ wave state with slightly-higher mass $D_0^*(j_q=1/2)$, both of which are expected to have wide widths of a few hundreds MeV.
} 
and $D_s$ meson mass spectra,
presently known and$/$or predicted through the above relations.

\section{Decay properties of $D_s$-mesons}

The observed properties of $D_s$-mesons to be examined are as follows\cite{rf4,rf5,rf6}:
\begin{eqnarray}
D_s(0^+;2.32) & \rightarrow & D_s(0^-;1.97) + \pi^0 \ \ \ {\rm observed}
, \nonumber\\
D_s(1^+;2.46) & \rightarrow & D_s(1^-;2.11) + \pi^0 \ \ \ {\rm observed}
. \label{eq18}\\
R(0^+) & \equiv & 
   \frac{ Br ( D_{s,0}^\chi (2.32) \rightarrow D_s^* \gamma ) }{Br ( D_{s,0}^\chi (2.32) \rightarrow D_s \pi^0 ) }
   < 0.078\ \ ({\rm CLEO}
)\ . \label{eq19}\\
R(1^+) & \equiv & 
   \frac{ Br ( D_{s,1}^\chi (2.46) \rightarrow D_s \gamma ) }{Br ( D_{s,1}^\chi (2.46) \rightarrow D_s^* \pi^0 ) }
   = 0.47\pm 0.10\ \ ({\rm Belle}
)\ . \label{eq20}\\
\Gamma_T [ D_s(0^+;&2.32&) ],\ \   \Gamma_T [ D_s(1^+;2.46) ] < 7{\rm MeV}. \label{eq21} 
\end{eqnarray}

{\it ($Pionic$ $transitions$)}\ \ \ \ The observed processes (\ref{eq18}) are iso-spin violating 
and considered to occur by the mixing 
of intermediate $\eta$ meson with $\pi$-meson.
From this picture and Eq.~(\ref{eq16}) we get the relation\cite{rf3,rf8},
\begin{eqnarray}
\Gamma ( D_s^+ (0^+) \rightarrow  D_s(0^-) \pi^0 ) &=&
\Gamma ( D_s^+ (1^+) \rightarrow  D_s(1^-) \pi^0 ) ,
\label{eq23}
\end{eqnarray}
which is consistent with the property (\ref{eq21}).
We can estimate phenomenologically the value of mixing parameter ${\rm sin}\theta$, by
using the experimental branching ratio\cite{rf7} of $D_n(c\bar n)$ meson to the iso-spin violating decay channel as
\begin{eqnarray}
({\rm sin}\theta )^2_{\rm exp}  &\approx& 
  \frac{ Br ( D_{s}^{*+}  \rightarrow D_s^+ \pi^0 ) (M_{D_s^*}^2 / q^3 ) }{ Br ( D_{s}^{*+}  \rightarrow D_s^+ \gamma ) } / 
  \frac{2 Br ( D^{*+}  \rightarrow D^+ \pi^0 ) (M_{D^*}^2 / q^3 ) }{ Br ( D^{*+} \rightarrow D^+ \gamma ) }   \nonumber\\ 
  &=& (0.9\pm 0.4)\cdot 10^{-3} .
\label{eq24}
\end{eqnarray}
This seems to be of reasonable order of magnitude as due to the virtual EM-interaction.
In order to estimate the absolute magnitude of the width (\ref{eq23}) in relation with those of the other HL-mesons,
we shall set up the chiral symmetric interaction Lagrangian
in the framework of covariant oscillator quark model (COQM)\cite{rfCOQM} as 
\begin{eqnarray}
{\cal S}_Y^I &=& \int d^4 x_1 d^4 x_2 {\cal L}(x_1,x_2) 
            \equiv \int d^4 X {\cal L}_I(X),\ {\cal L}={\cal L}^{ND}+{\cal L}^{AX}
\label{eq25}\\
{\cal L}^{ND} &=& g_{ND} \langle \Phi (x_1,x_2) M(x_2) \bar\Phi (x_1,x_2) \rangle  .
\nonumber\\
{\cal L}^{AX} &=& g_{AX} \langle \Phi (x_1,x_2) 
    ( (\stackrel{\rightarrow}{\partial_{2,\mu}}+\stackrel{\leftarrow}{\partial_{2,\mu}})  
      + i\sigma_{\mu\nu}(\stackrel{\rightarrow}{\partial_{2,\nu}}-\stackrel{\leftarrow}{\partial_{2,\nu}})  )
                              [ \partial_{2,\mu} M(x_2) ] \bar\Phi (x_1,x_2) \rangle  .
\nonumber\\
M &\equiv& s - i \gamma_5 \phi\ ( s\equiv s^a \lambda^a /\sqrt 2,\   \phi\equiv \phi^a \lambda^a /\sqrt 2), \nonumber\\
\Phi & \propto & (1-iv\cdot\gamma )
( i \gamma_5 D + i \tilde\gamma_\nu D_\nu^* + D_0^\chi + i \gamma_5 \tilde\gamma_\nu D_\nu^{*\chi}  ), \nonumber\\
\bar\Phi &\equiv& \gamma_4 \Phi^\dagger \gamma_4,\ \   
D=(\sqrt{2M_{D^0}}D^0,\sqrt{2M_{D^+}}D^+,\sqrt{2M_{D_s}}D_s^+)\ etc.\ ,\label{eq28}
\end{eqnarray}
where only the Yukawa-type interaction of the scalar $(s)$ and pseudo-scalar $(\phi )$ nonets 
with the light quarks in the HL-meson is taken into account.

The interaction (\ref{eq25}) consists of the two terms:\\
Firstly the $g_{ND}$ term (Yukawa interaction in non-derivative form) 
gives dominant (compared to the $g_{AX}$ term) contribution 
to the (quark-) spin non-flip processes. 
In spontaneous breaking of chiral symmetry,
 $s(=s^a\lambda^a/\sqrt 2)$ takes the vacuum expectation value
which induces the mass-splittings between chiral partners through
the equation $\Delta M^\chi (c\bar n)=2g_{ND}a$ and 
$\Delta M^\chi (c\bar s)=2g_{ND}b$ (, see Eq.~(\ref{eq16})),
as explained in \S 2.\\
Secondly the $g_{AX}$ term in the interaction (\ref{eq25}) corresponds to the extended
PCAC term, and concerns dominantly (compared to the $g_{ND}$ term) to the spin-flip processes.
The formula of the relevant pionic decay amplitudes and widths are given in Table \ref{tab1}. 

\begin{table}
\begin{tabular}{l|l}
\hline
processes$/$widths & invarinat amplitude ${\cal A} (\equiv \frac{\cal M}{\sqrt{MM^\prime}})/$helicity 
amplitude ${\cal A}_{h_f,h_i}$ \\
\hline
$D^{*+}(\epsilon_\mu (P))\rightarrow D^0(P^\prime )\pi^+(q)$  &  
   ${\cal A}_{h_{D^*}}= - \{ g_{AX} ( -q^2 + \frac{2m_2}{m_1+m_2}MM^\prime (\omega +1) )+g_{ND}  \} \epsilon^{(h_{D^*})}\cdot v^\prime $\\ 
       \multicolumn{1}{r|}{$\Gamma = \frac{M_D |{\bf q}|}{8\pi M_{D^*}} \frac{1}{3} |{\cal A}_0|^2$ } 
   &  ${\cal A}_{0}= - \{ g_{AX} ( -q^2 + \frac{2m_2}{m_1+m_2}MM^\prime (\omega +1) )+g_{ND}  \} \omega_3$\\  
\hline
$D^{\chi}_{n(s)0}(P)\rightarrow D_{n(s)}(P^\prime ) \pi (q) $  &  
   ${\cal A}= - \left(  g_{ND} +  g_{AX} ( -q^2 + \frac{2m_2}{m_1+m_2}MM^\prime (\omega -1) )  \right) (\omega +1) $\\ 
\multicolumn{2}{c}{ $\Gamma_{D_{n0}^\chi \rightarrow D_n \pi} 
                   = \frac{M_D |{\bf q}|}{8\pi M_{D^\chi_0}} \frac{3}{2} |{\cal A}|^2$,
                    $\Gamma_{D_{s0}^{\chi +} \rightarrow D_s^+ + \pi^0} 
                   = \frac{M_D |{\bf q}|}{8\pi M_{D^\chi_0}}|{\cal A}|^2 {\rm sin}^2\theta_{\pi^0} $
         \ \ \ \ \ \ \ \ \ \ \ \ \ \ \ \ \ \ \ \ \ \ \ \ \ \ \ \ \ \ \ \  } \\  
\hline
$D^{\chi}_{n(s)1}(\epsilon_\mu (P)) \rightarrow D_{n(s)}^* (\epsilon_\mu (P^\prime)) \pi (q) $  &  
   ${\cal A}_{h_{D^*},h_{D_1^\chi}}=$\\
     &  $ i  (  g_{ND} -q^2  g_{AX} ) ( \epsilon^{(h_{D_1^\chi)}}\cdot\tilde\epsilon^{(h_{D^*})} (\omega +1) 
                                            + \epsilon^{(h_{D_1^\chi)}}\cdot v^\prime \tilde\epsilon^{(h_{D^*})}\cdot v  )$\\
     &  $+ i g_{AX}   \frac{2m_2}{m_1+m_2}MM^\prime   ( \epsilon^{(h_{D_1^\chi)}}\cdot\tilde\epsilon^{(h_{D^*})} (\omega^2 -1) 
                                            + \omega \epsilon^{(h_{D_1^\chi)}}\cdot v^\prime \tilde\epsilon^{(h_{D^*})}\cdot v  )$\\
\multicolumn{1}{r|}{$\Gamma = \frac{M_D |{\bf q}|}{16\pi M_{D^\chi_1}}   
     \sum_{ h_{D_1^\chi} } |{\cal A}_{h_{D_1^\chi}, h_{D^*} } |^2$ } 
   & ${\cal A}_{++}={\cal A}_{--}
      =i \left(  g_{ND} +  g_{AX} ( -q^2 + \frac{2m_2}{m_1+m_2}MM^\prime (\omega -1) )  \right) (\omega +1)  $ \\  
     & ${\cal A}_{00} = i \left(  g_{ND}  - q^2 g_{AX}  \right)  (\omega +1)  $ \\
\hline
\end{tabular}
\caption{Formula of pionic decay amplitudes and widths.
The invariant(helicity) ampliudes for respective processes are given first(second).
The quantities in the columns are defined as; 
$M(M^\prime)$ is the mass of the initial(final) $D_{n(s)}$ meson; $q_\mu \equiv P_\mu -P^\prime_\mu$;
$|{\bf q}|=M^\prime \omega_3\ (M^\prime \omega)$ is momentum (energy) of the final $D_{n(s)}$ meson
in the CM system;
$m_1(m_2)$ constituent quark(antiquark) mass.
}
\label{tab1}
\end{table}

The $g_{ND}$ is fixed, from the experimental value of $\Delta M^\chi (c\bar s)=2g_{ND}b=350$MeV,
with the value $g_{ND}=1.848$, while the $g_{AX}$ is fixed, from the experimental decay width 
$\Gamma (D^{*+}\rightarrow D^0 \pi^+ ) = (96\pm 23)\times 0.68$keV, with $g_{AX}=5.10$GeV$^{-2}$. 
By using these values, we can predict the absolute values of the relevant pionic decay widths as
\begin{eqnarray}
\Gamma ( D_{n,0}^\chi \rightarrow  D_n \pi ) &=& 
   \Gamma ( D_{n,1}^\chi \rightarrow  D_n^* \pi ) = 158{\rm MeV}\ ,  \label{eq29}\\
\Gamma ( D_{s,0}^\chi \rightarrow  D_s \pi^0 ) &=& 
   \Gamma ( D_{s,1}^\chi \rightarrow  D_s^* \pi^0 ) = 155\pm 70{\rm keV}\ ,  
\label{eq30}
\end{eqnarray}
where, in deriving Eq.~(\ref{eq30}), the estimated value Eq.~(\ref{eq24}) is used.
The value of width (\ref{eq30}) is consistent with the experiment (\ref{eq21}).\\

{\it ($Radiative$ $decay$)}\ \ \ \ In order to treat systematically 
all the radiative transitions between the HL-mesons
we shall set up the basic EM-interaction Lagrangian in the framework of COQM\cite{rfCOQM}, as 
\begin{eqnarray}
{\cal S}_I^{EM} &=& \int d^4x_1 d^4x_2 \sum_{i=1,2} j_{i,\mu} (x_1,x_2) A_\mu (x_i) 
  = \int d^4 X \sum_i J_{i,\mu}(X)A_\mu (X),\nonumber\\
j_{i,\mu} (x_1,x_2) &=& -i e_i \left( (m_1+m_2)/m_i\right)  \langle  \bar\Phi_U 
   ( \partial_{i\mu}^- + i g_M \sigma_{\mu\nu}^{(i)} \partial_{i,\nu}^+ ) \Phi_U  \rangle ,\ \ \ \ \ 
\label{eq31}\\
\Phi_U  &\equiv& \Phi_U (-i\stackrel{\leftarrow}{v}\cdot\gamma ),\ \   
   \bar\Phi_U  \equiv \bar\Phi_U (-i\stackrel{\leftarrow}{v}\cdot\gamma ),\nonumber
\end{eqnarray}
where $\Phi_U$ is the unitary correspondent of $\Phi$, so defined as 
$\langle \bar\Phi_U \Phi_U  \rangle \rightarrow  \langle \Phi^\dagger \Phi  \rangle $
at the rest frame. Here it is to be noted that our effective current $J_{i,\mu}(X)$ is obtained
through the ``minimal substitution" of $(\partial_{i,\mu}\rightarrow \partial_{i,\mu} -i e_i A_\mu (x_i))$,
and accordingly it is conserved in the ideal limit.

Our effective current has also another remarkable feature due to the covariant nature of our scheme.
The spin-current interaction (the second term in Eq.~(\ref{eq31})) leads to the Hamiltonian
\begin{eqnarray}
{\cal H}^{(i)\ spin} &\equiv&  J_{\mu}^{(i) spin} A_\mu = 
\mu^{(i)} \mbox{\boldmath$\sigma$}^{(i)}\cdot \mbox{\boldmath$B$}
+ d^{(i)} \rho_1^{(i)} \mbox{\boldmath$\sigma$}^{(i)}\cdot \mbox{\boldmath$E$},
\ \ \mu^{(i)}=d^{(i)}=e_i/2m_i .\ \ \ \ 
\label{eq32}
\end{eqnarray}
This shows that our Hamiltonian contains the interaction through
the ``intrinsic electric dipole" $d\rho_1\mbox{\boldmath$\sigma$}$ as well as the one through
the magnetic dipole $\mu \mbox{\boldmath$\sigma$}$.
The ``intrinsic dipole" gives contributions only for the transitions between chiralons and Paulons, 
while does none for the other transitions.

From the effective currents $J_{i\mu}$ in Eq.~(\ref{eq31}), we can derive 
the formula of the relevant radiative decay amplitudes and widths, which are given in Table \ref{tab2}.

\begin{table}
\begin{tabular}{l|l}
\hline
processes$/$widths & invariant amplitude ${\cal M}/$helicity amplitude ${\cal M}_{h_\gamma , h_D}$ \\
\hline
$D_{d,s}^{*+}(\epsilon_\mu (P)) \rightarrow D_{d,s}^+(P^\prime ) \gamma (\eta_\mu (q))$
   &  ${\cal M}_{h_\gamma ,h_{D^*}}=\frac{d}{2}\left( \frac{e_1}{2m_1}+\frac{e_2}{2m_2} \right) (M+M^\prime )
         \frac{1}{i}\epsilon_{\mu\nu\alpha\beta} \tilde\eta_\mu^{(h_\gamma )} 
                  \epsilon_\nu^{(h_{D^*})} v_\alpha^\prime v_\beta$ \\ 
\multicolumn{1}{r|}{$\Gamma = \frac{\alpha |{\bf q}|}{2M_{D^*_{d,s}}^2} \frac{2}{3} |{\cal M}_{-,+}|^2$} 
   &  ${\cal M}_{-,+}=-{\cal M}_{+,-}=-i\frac{d}{2}\left( \frac{e_1}{2m_1}+\frac{e_2}{2m_2} \right) (M+M^\prime )\omega_3$  \\  
\hline
$D_{s0}^{\chi}(P) \rightarrow D_{s}^*(\epsilon_\mu (P^\prime )) \gamma (\eta_\mu (q))$
   &  ${\cal M}_{h_\gamma ,h_{D_s^*}}=\frac{d}{2} \tilde\eta^{(h_\gamma )}\cdot\tilde\epsilon^{h_{D_s^*}} \frac{M^2-M^{\prime 2} }{2MM^\prime}  
     \left(  \frac{e_1}{2m_1}(M-M^\prime )  -\frac{e_2}{2m_2} (M+M^\prime )  \right)$\\  
\multicolumn{1}{r|}{$\Gamma = \frac{\alpha |{\bf q}|}{M_{D^\chi_{s0}}^2} |{\cal M}_{+,+}|^2$} 
   &  ${\cal M}_{+,+}={\cal M}_{-,-}=\frac{d}{2} \frac{M^2-M^{\prime 2} }{2MM^\prime}  
     \left( - \frac{e_1}{2m_1}(M-M^\prime )  + \frac{e_2}{2m_2} (M+M^\prime )  \right)$\\  
\hline
$D_{s1}^{\chi}(\epsilon_\mu (P)) \rightarrow D_{s}(P^\prime ) \gamma (\eta_\mu (q))$
   &  ${\cal M}_{h_\gamma ,h_{D_{s1}^\chi}}=i\frac{d}{2} \tilde\eta\cdot\epsilon \frac{M^2-M^{\prime 2} }{2MM^\prime}  
     \left(  \frac{e_1}{2m_1}(M-M^\prime )  -\frac{e_2}{2m_2} (M+M^\prime )  \right)$\\  
\multicolumn{1}{r|}{$\Gamma = \frac{\alpha |{\bf q}|}{2M_{D^\chi_{s1}}^2} \frac{2}{3} |{\cal M}_{-,+}|^2$} 
   &  ${\cal M}_{-,+}={\cal M}_{+,-}=i\frac{d}{2} \frac{M^2-M^{\prime 2} }{2MM^\prime}  
     \left( \frac{e_1}{2m_1}(M-M^\prime )  - \frac{e_2}{2m_2} (M+M^\prime )  \right)$\\  
\hline
$D_{s1}^{\chi}(\epsilon_\mu (P)) \rightarrow D_{s0}^\chi (P^\prime ) \gamma (\eta_\mu (q))$
   &  ${\cal M}_{h_\gamma ,h_{D_{s1}^\chi}}=-\frac{d}{2}\left( \frac{e_1}{2m_1}+\frac{e_2}{2m_2} \right) (M+M^\prime )
         \epsilon_{\mu\nu\alpha\beta} \tilde\eta_\mu \epsilon_\nu v_\alpha^\prime v_\beta$ \\ 
\multicolumn{1}{r|}{$\Gamma = \frac{\alpha |{\bf q}|}{2M_{D^\chi_{s1}}^2} \frac{2}{3} |{\cal M}_{-,+}|^2$} 
   &  ${\cal M}_{-,+}=-{\cal M}_{+,-}=-\frac{d}{2} 
     \left( \frac{e_1}{2m_1} + \frac{e_2}{2m_2} (M+M^\prime ) \omega_3 \right)$\\  
\hline
\end{tabular}
\caption{Formula of radiative decay amplitudes and widths. 
$e_1$($-e_2$) is the charge of the first(second) constituent.
In the case of $D_s^+=c\bar s$, $e_1(e_2)=\frac{2}{3}e(-\frac{1}{3}e) \cdots$.
$d\equiv 2(m_1+m_2)$.
For other quantities, see, the caption in Table \ref{tab1}. 
}
\label{tab2}
\end{table}

By using Table \ref{tab2} we can predict the widths 
for all the radiative spin-flip transitions between ground state $D_s$ mesons.
The results are given in Table \ref{tab3}, where for reference, the width 
for the transition of $D_n$ meson, $D_n^{*,+}(1^-) \rightarrow D_n^+(0^-) \gamma$
is given in comparison with experiments. 
There, we have also shown the predicted values by the other chiral model. 

\begin{table}
\begin{tabular}{clc|c|c}
\hline
   & Processes & $P/\chi \rightarrow P/\chi$ & $\Gamma$(keV) & $\Gamma$(keV) \\
    &                  &                                         &  ours   &  others \cite{rf9} \\
\hline
(a) & $D_s(1^-)\rightarrow D_s(0^-)\gamma$ & $P\rightarrow P$ & 0.33 & 0.43 \\
(b) & $D_s(1^+)\rightarrow D_s(0^+)\gamma$ & $\chi\rightarrow \chi$ & 0.26 & 0.43 \\
(c) & $D_s(0^+)\rightarrow D_s(1^-)\gamma$ & $\chi\rightarrow P$ & 21 & 1.74 \\
(d) & $D_s(1^+)\rightarrow D_s(0^-)\gamma$ & $\chi\rightarrow P$ & 93 & 5.08 \\
\hline
   &   $D_n^{*,+}(1^-)\rightarrow D_n^+(0^-)\gamma$ &  $P\rightarrow P$ & 
    \multicolumn{2}{c}{$1.14({\rm theor})\leftrightarrow  1.54\pm 0.53({\rm exp})$} \\
\hline
\end{tabular}
\caption{$\gamma$-decay widths for spin-flip transitions between the ground state $D_s$ mesons.
$\alpha =1/137.036$. The constituent quark masses are fixed with the values,
$m_u=m_d \equiv m_n = M_\rho /2$, $m_s=M_\phi /2$, $m_c=M_{J/\psi}/2$.
}
\label{tab3}
\end{table}

From the results in Table \ref{tab3} we see that our model gives the much 
larger widths for the transitions, (c) and (d),
from chiralons to Paulons, compared to the other chiral model (, reflecting the above mentioned feature (\ref{eq32})
of our currents,) while does the width of almost the same amount for transitions, (a) (and (b)), 
from Paulons(chiralons) to Paulons(chiralons).
This difference is considered to come from the different identification of the relevant mesons
in the two cases:
The narrow $D_s$ mesons are assigned as the conventional $P$-wave excited states 
in the other model\cite{rf9},
while they are the $S$-wave chiral states other than the $P$-wave Pauli-states in our scheme.

{\it ($Branching$ $ratios$ $between$ $radiative$ $to$ $pionic$ $decay$ $widths$)}\ \ \ \ 
From the predicted values of pionic (Eq.~(\ref{eq30})) and radiative (Table \ref{tab3}) decay widths we obtain
the ratios between them as follows:
\begin{eqnarray}
R(0^+) &=& 0.14\stackrel{+0.11}{\scriptstyle -0.05} ,\ \ R(1^+) = 0.6\stackrel{+0.5}{\scriptstyle -0.2} ,
\label{eq33}
\end{eqnarray}
which seems to be consistent with the experiments 
Eqs.~(\ref{eq19}) and (\ref{eq20}).

\section{Concluding Remarks}

\renewcommand{\labelitemi}{$\circ$}
\begin{itemize}

\item  The $D_s(2317)$ and $D_s(2463)$ mesons are shown consistently assigned as the chiralons with 
$J^P=0^+$ and $1^+$ in the $(c\bar s)$ ground states. 

\item The decay width of $(D_n^\chi (1^+) \rightarrow D_n^* (1^-) \pi )$ is predicted as
      $\Gamma \simeq 130$MeV, and the radiative decay widths of chiral states into Pauli-states 
      are predicted to be remarkably larger than those estimated in other works. 
      These are to be checked experimentally.

\item Further experimental search for chiralons is desirable.

\end{itemize}

\end{document}